\documentclass[aps,prl,superscriptaddress,floatfix,twocolumn,longbibliography]{revtex4-2}
\usepackage{amssymb,amsmath,amsfonts}
\usepackage{graphicx}
\usepackage{theorem}
\usepackage{dcolumn}
\usepackage{braket}
\usepackage{bm}
\usepackage{setspace}
\usepackage{eurosym}
\usepackage{amsfonts,amsmath}
\usepackage{mathrsfs}
\usepackage{amssymb}
\usepackage{graphicx}
\usepackage{bm}
\usepackage{siunitx}
\usepackage{orcidlink}

\begin{document}

\title{Electron Lateral Trapping Induced by Non-Uniform Thickness in Solid
Neon Layers}
\author{Toshiaki Kanai~\orcidlink{0000-0002-9446-2497}}
\author{Chuanwei Zhang}
\email{Email: chuanwei.zhang@wustl.edu}
\affiliation{Department of Physics, Washington University, St.\,Louis,
Missouri 63130, USA}
\begin{abstract}
Recent experimental advances highlight electron charge qubits floating above
solid neon as an emerging promising platform for quantum computing, but the
physical origin of single-electron lateral trapping is still not fully
understood. While prior theoretical work has mainly examined electrons above
bulk solid neon, experimental systems usually feature neon layers of only $\lesssim 10$ nm 
thickness and non-uniformity, highlighting unresolved
questions about how thickness influences electron trapping. Here we
theoretically investigate the effect of finite thickness and non-uniformity of solid
neon layers on electron trapping. For a 10 nm layer, the electron
binding energy is enhanced threefold compared to bulk.
Exploiting this thickness dependence, we propose a nanopatterned-substrate
mechanism in which engineered thickness variations generate lateral trapping
potentials for electrons. The lateral trapping potential can be finely tuned
by a perpendicular electric field. Such non-uniform-thickness induced
electron charge qubits open a viable pathway toward building multi-qubit
systems for quantum computation.
\end{abstract}

\maketitle

%\email{Email: chuanwei.zhang@wustl.edu}
%

Semiconductor-based electron charge qubits have emerged as a leading
hardware platform for quantum computation, benefiting from compatibility
with modern microelectronics and advanced fabrication technologies~\cite{chatterjee_semiconductor_2021, heinrich_quantum-coherent_2021,
stano_review_2022}. However,
their performance is limited by short coherence times arising from charge
noise~\cite{petersson_quantum_2010} and phonon~\cite{ranni_decoherence_2024}. Recent 
theoretical and experimental advances demonstrate
that a single-electron qubit floating above an ultra-clean solid neon
surface (e-Ne) exhibits dramatically reduced charge noise and significantly
longer coherence times \cite{zhou_single_2022, zhou_electron_2024,
li_noise-resilient_2025, li_electron_2025}.
Combined with recent progress in controlling
interactions between two electrons through cross resonance~\cite{li_coherent_2025}, e-Ne qubits are
emerging as a promising platform for solid-state quantum computing, and advanced methods utilizing spin degrees of freedom are also being explored~\cite{xie_high-fidelity_2024,tian_nbtin_2025, pan_nonlinear_2025, kanagin_impurities_2025}.
However, a major challenge persists:
the fundamental mechanism responsible for electron trapping remains poorly understood, impeding 
efforts toward reproducible single-qubit design, precise control and readout, 
and the construction of scalable multi-qubit architectures.

\begin{figure}[t]
\centering
\includegraphics[width=0.95\linewidth]{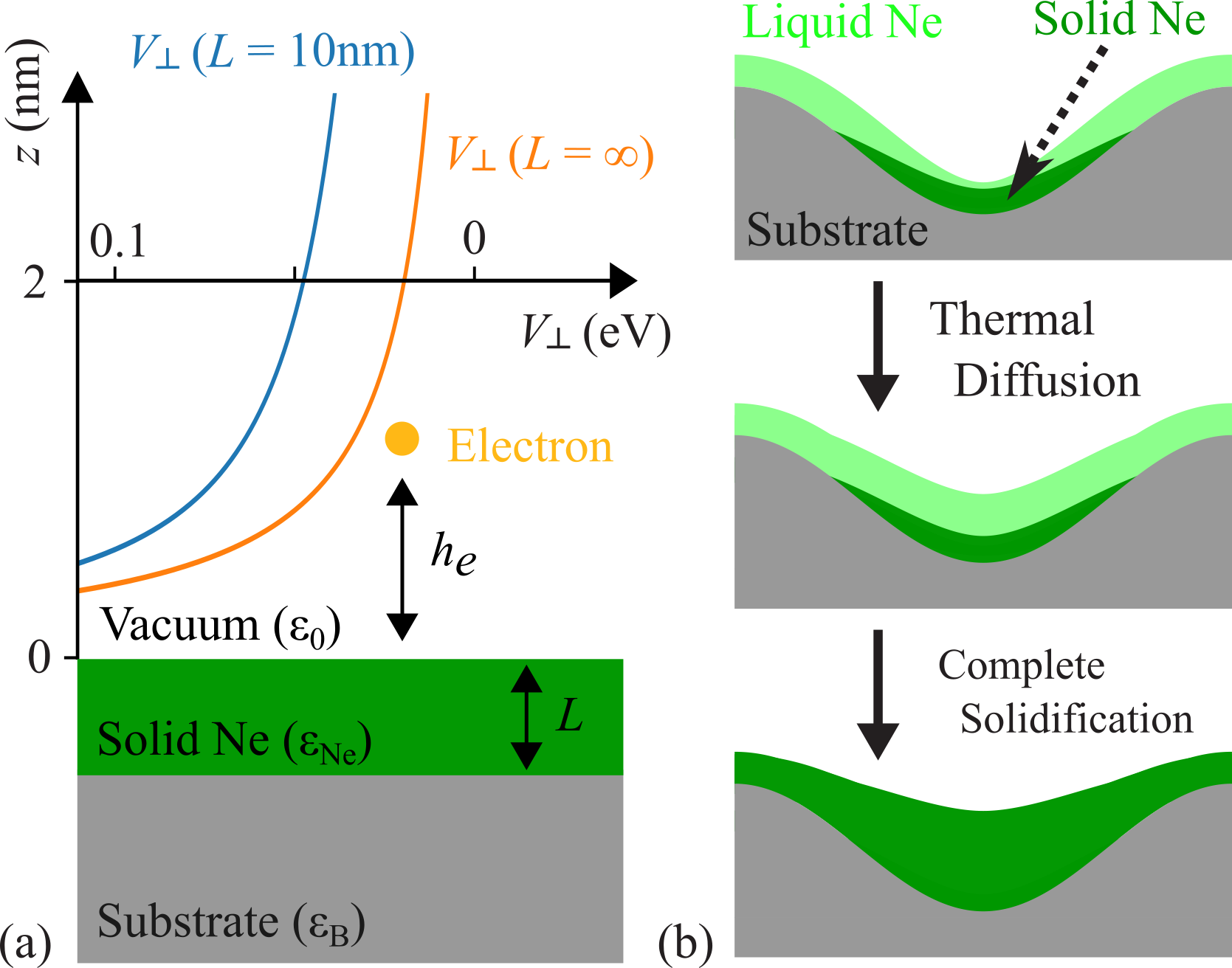}
\caption{(a) Schematic illustration of electrons hovering above a solid Ne layer. 
The blue and orange curves represent the perpendicular potential $V_{\perp}$ for a finite thickness
($L = 10~\mathrm{nm}$) and for the bulk case ($L = \infty$), respectively.
(b) Schematic illustration of the formation process of a non-uniform solid Ne layer.
 Initially, a uniform liquid Ne film forms above the substrate. Due to the Gibbs–Thomson effect, 
 solidification begins preferentially at the bottom of surface depressions. 
 The remaining liquid then thermally diffuses to re-establish a uniform layer above the newly solidified regions. 
 The interplay of these processes ultimately leads to a non-uniform solid Ne layer.
}
\label{fig:2D_schematic}
\end{figure}

Figure~\ref{fig:2D_schematic}(a) illustrates the e-Ne qubit platform, in
which a solid neon layer is deposited on a substrate such as niobium (Nb) or
silicon (Si). An excess electron above the flat neon surface induces a
positive image charge that attracts it back toward the surface. This
attraction, together with the Pauli-exclusion barrier from the outer-shell
electrons of neon atoms \cite{cole_image-potential-induced_1969,
cole_electronic_1971, jin_quantum_2020}, creates a deep trapping potential
in the direction perpendicular to the neon surface. The main challenge lies in understanding the
mechanism of lateral confinement. A natural intuition is that the
electrostatic potentials generated by surrounding electrodes should
define the lateral trapping; however, this perspective disagrees with experimental observations.
For instance, the observed shifts in the excitation spectrum due to the reduced electrostatic 
potentials are much
smaller than expected, and electrons remain bound to the neon surface even
in the absence of such potentials. This contradiction has
motivated the proposal of a new lateral trapping mechanism arising from
surface geometry, specifically Gaussian-shaped bumps or valleys \cite%
{kanai_single-electron_2024}. 

However, all existing theoretical models have
assumed bulk neon, whereas in experiments the neon layer is only $\sim 5-10$ nm thick. 
Moreover, the solid neon layer often exhibits non-uniform
thickness due to substrate morphology, as evidenced by recent experiments
showing lateral electron trapping above neon layers formed on rough Si
substrates \cite{zheng_surface-morphology-assisted_2025}. This
theoretical-experimental discrepancy raises two key questions: (i) how does
neon layer thickness affect the perpendicular and lateral trapping potentials;
and (ii) can patterned substrates be used to engineer neon layer thickness
for controlled lateral trapping and reproducible design of single-electron charge qubits?

In this Letter, we address these two key questions by investigating 
how the finite thickness and spatial non-uniformity of the solid neon layer 
shape the trapping potential of an excess electron. 
Our main findings are:

i) The binding energy for the electron above a $10~\mathrm{nm}$-thick solid Ne 
layer reaches $44.6~\mathrm{meV}$ on a superconducting substrate and 
$40.0~\mathrm{meV}$ on a silicon substrate, both markedly exceeding the 
bulk value of $15.7~\mathrm{meV}$.

ii) The thickness variations on the order of $\Delta L \sim 3~\mathrm{nm}$ 
can induce ground-state energy shifts of approximately $10~\mathrm{meV}$ 
for the excess electron, revealing a fundamental lateral trapping mechanism arising from thickness 
non-uniformity. Building on this mechanism, we propose a single-electron charge qubit 
confined by the potential created by an engineered nanopillar on the flat substrate, paving the 
way toward scalable multi-qubit architectures based on patterned nanopillar arrays.

iii) The lateral trapping potential and electron excitation energy
can be finely tuned through a uniform electric field applied in the perpendicular direction. 
The lateral excitation energy has a nonlinear dependence on the direction of the applied field.

\paragraph{Non-uniform neon layer thickness induced by substrate inhomogeneity. }

In a typical experimental procedure~\cite{zhou_single_2022}, 
liquid Ne is introduced into a cryogenic cell and gradually cooled 
to approximately $10~\mathrm{mK}$, forming a solid layer with a thickness of $5$–$10~\mathrm{nm}$ 
on the substrate. A recent study~\cite{zheng_surface-morphology-assisted_2025} 
reported numerous surface valleys on silicon substrates, 
with widths of about $200~\mathrm{nm}$ and depths of roughly $25~\mathrm{nm}$, dimensions significantly 
larger than the typical thickness of the solid Ne layer. The gravitational potential difference associated with 
such height variations is on the order of $5\times10^{-11}~\mathrm{meV}$, which is negligible compared 
to the Ne–Ne interaction energy ($3.1~\mathrm{meV}$ or $\sim36~\mathrm{K}$) and to thermal fluctuations. 
As a result, a uniform liquid layer is expected to form initially. 
However, the substrate roughness can strongly influence the solidification process near the triple 
point~\cite{esztermann_triple-point_2002}, locally shifting the melting temperature via the Gibbs–Thomson effect~\cite{galenko_phase_2024,glicksman_principles_2011}, and thereby introducing thickness non-uniformity 
in the solid layer. The shift in melting temperature, denoted $\Delta T_{R}$, is governed by 
the Gibbs–Thomson relation:
\begin{equation}
\Delta T_{R}=-\frac{2\gamma _{\mathrm{SL}}V_{m}}{r_{c}\Delta H_{f}}T_{%
\mathrm{bulk}}
\label{eq:GT}
\end{equation}%
where $\gamma_{\mathrm{SL}}$ is the solid–liquid interfacial energy 
(estimated to be $\gamma_{\mathrm{SL}} \approx 4.36~\mathrm{mJ/m^{2}}$ 
for neon~\cite{hansen_ground_1968,broughton_molecular_1986}), 
$V_{m} \approx 13.98\times10^{-6}~\mathrm{m^{3}/mol}$ is the molar volume of solid neon, 
$r_{c}$ is the local radius of curvature, 
$\Delta H_{f} = 328~\mathrm{J/mol}$~\cite{rumble_crc_2025} is the molar enthalpy of fusion, 
and $T_{\mathrm{bulk}} = 24.56~\mathrm{K}$~\cite{rumble_crc_2025} is the bulk melting temperature. 
This leads to an approximate relation $\Delta T_{R} \approx -9.12~[\mathrm{K\cdot nm}]/r_{c}$.
For typical curvature values of $r_{c} = \pm 10~\mathrm{nm}$ ($+$ and $-$ correspond to bump and valley, respectively), the melting-temperature shift reaches 
 $\Delta T_{R} \approx \mp 0.9~\mathrm{K}$, with the sign of $\Delta T_{R}$ determined by the curvature direction.
 Consequently, solidification tends to initiate preferentially at the bottom of surface depressions. 

Neon atoms in the liquid phase undergo thermal diffusion, characterized by a self-diffusion
coefficient on the order of $10^{-3}~\mathrm{mm^{2}/s}$ at $25~\mathrm{K}$~\cite{beilogua_coefficient_1971}, 
allowing diffusion over distances of roughly $100~\mathrm{nm}$ within about $10~\mu\mathrm{s}$. 
The combined effects of partial solidification and rapid thermal diffusion, 
both occurring much faster than the typical cooling rate of $\sim1~\mathrm{K/h}$, 
lead to preferential accumulation of Ne atoms in surface depressions. 
As a result, thicker solid layers gradually form in these regions, 
as illustrated in Fig.~\ref{fig:2D_schematic}(b). 
It is also worth noting that mechanical vibrations during rapid cooling 
may contribute to additional thickness fluctuations in the actual system.

\begin{figure*}[t]
\centering
\centering
\includegraphics[width=0.98\linewidth]{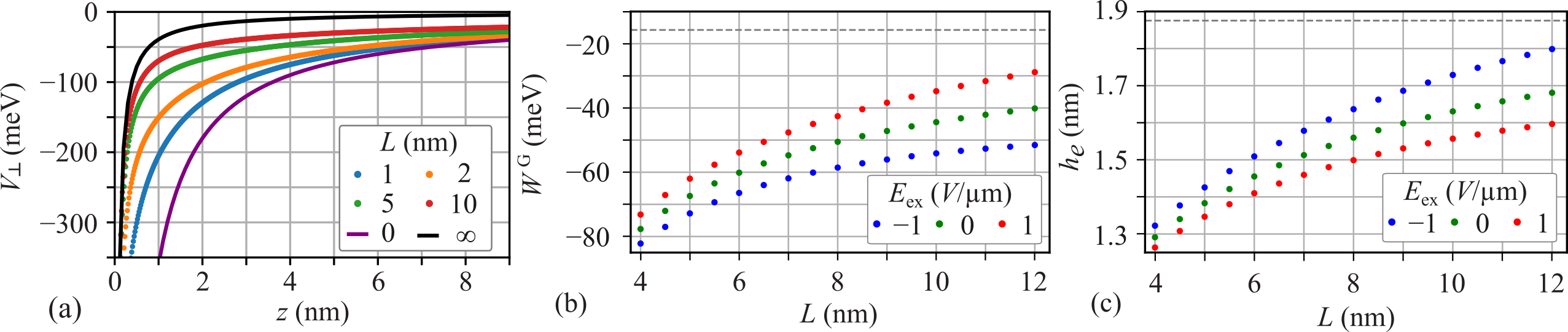}
\caption{Perpendicular potential profile $V_L^z(z)$ for various Ne layer thicknesses $L$. 
(b) Ground-state energy $W_L^{\mathrm{G}}$ and (c) mean electron height $h_e$ as 
functions of $L$ under different external electric fields $E_{\mathrm{ex}}$. 
The dashed lines indicate the corresponding values in the bulk limit ($L \to \infty$).}
\label{fig:perpendicular}
\end{figure*}

\paragraph*{Thickness dependence of perpendicular trapping potential.}

We begin by analyzing the trapping potential experienced by an excess
electron in the direction perpendicular to a flat solid neon surface (Fig. \ref{fig:2D_schematic}(a)).
The system consists of a vacuum in the upper half-space ($z > 0$), 
a solid Ne layer of thickness $L$ occupying the region $-L < z < 0$, 
and a substrate below ($z < -L$). The spatially dependent dielectric permittivity is defined as:
\begin{equation}
\varepsilon(L,z) = 
\begin{cases}
\varepsilon_0 & \mathrm{for} \hspace{4pt} 0 < z \\ 
\varepsilon_\mathrm{Ne} & \mathrm{for} \hspace{4pt} -L < z < 0 \\ 
\varepsilon_\mathrm{B}, & \mathrm{for} \hspace{4pt} z < -L%
\end{cases}
\label{eq:DielectricConst}
\end{equation}
where $\varepsilon_0$, $\varepsilon_{\mathrm{Ne}} = 1.244\,\varepsilon_0$, 
and $\varepsilon_{\mathrm{B}}$ are the permittivity of vacuum, solid neon and substrate, respectively. 
Assuming a point charge $-e$ located at $\vec{r}_0 = (\rho = 0, z = z_0)$ in cylindrical coordinates, 
the electrostatic potential satisfies the Poisson equation~\cite{jackson_classical_2009}:
\begin{eqnarray}
\nabla \cdot \left(\varepsilon(r) \nabla\phi(\vec{r})\right) &=& -\rho_e(%
\vec{r}) \\
\rho_e(\vec{r}) &=& -e \delta(\vec{r} - \vec{r}_0).
\end{eqnarray}

The perpendicular trapping potential $V_\perp(L, z)$ experienced by the electron due to the imaging charge can be well approximated \cite{sup}
by evaluating the potential at $\vec{r}_0 = \vec{r}$~\cite{cole_image-potential-induced_1969}:
\begin{eqnarray}
V_\perp(L, z) &=& -\frac{\phi(\vec{r})}{2} \\
&=& \frac{e^2}{8 \pi \varepsilon_0} \int_0^\infty 
\Lambda_L(k) e^{-2kz} dk,
\end{eqnarray}
where the self-interaction term is subtracted, and $\Lambda_L$ is the
reflection coefficient: 
\begin{eqnarray}
\Lambda_L(k) &=& \frac{(\varepsilon_{0} - \varepsilon_{B}) \varepsilon_{Ne}
+ (\varepsilon_{0} \varepsilon_{B} - \varepsilon_{Ne}^2) \tanh(kL)}{%
(\varepsilon_{0} + \varepsilon_{B}) \varepsilon_{Ne} + (\varepsilon_{0}
\varepsilon_{B} + \varepsilon_{Ne}^2) \tanh(kL)}.
\end{eqnarray}
For a superconducting substrate ($\varepsilon_{\mathrm{B}} \to \infty$), this expression simplifies to:
\begin{eqnarray}
\Lambda_L(k) &=& \frac{\varepsilon_0 \tanh(kL) - \varepsilon_{Ne}}{%
\varepsilon_0 \tanh(kL) + \varepsilon_{Ne}}.
\end{eqnarray}

In the bulk limit ($L\to \infty$), the potential reduces to the well-known
image charge potential above bulk solid Ne:
\begin{eqnarray}
V_\perp(L=\infty, z) &=& -\frac{1}{8\pi \varepsilon_0} \frac{\varepsilon_%
\mathrm{Ne} - \varepsilon_0}{\varepsilon_\mathrm{Ne} + \varepsilon_0} \frac{%
e^2}{2z},
\end{eqnarray}
whereas in the limit of vanishing thickness ($L \to 0$), the potential reduces to that produced by a mirror charge $e$:
\begin{eqnarray}
V_\perp(L=0, z) &=& - \frac{1}{8\pi \varepsilon_0} \frac{e^2}{2z}.
\end{eqnarray}

Figure~\ref{fig:perpendicular}(a) shows $V_\perp$ 
for various solid neon layer thicknesses $L$, revealing a deeper trapping 
potential for thinner layers. In all cases, the potential asymptotically 
approaches zero as $z \to \infty$. Although an ideal superconducting substrate 
is assumed in this study, the analysis also applies to semiconductors with high 
permittivity, such as Si ($\varepsilon \approx 12\,\varepsilon_0$) and 
GaAs ($\varepsilon \approx 13\,\varepsilon_0$)~\cite{rumble_crc_2025}. 

\begin{figure*}[t]
\centering
\includegraphics[width=0.99\linewidth]{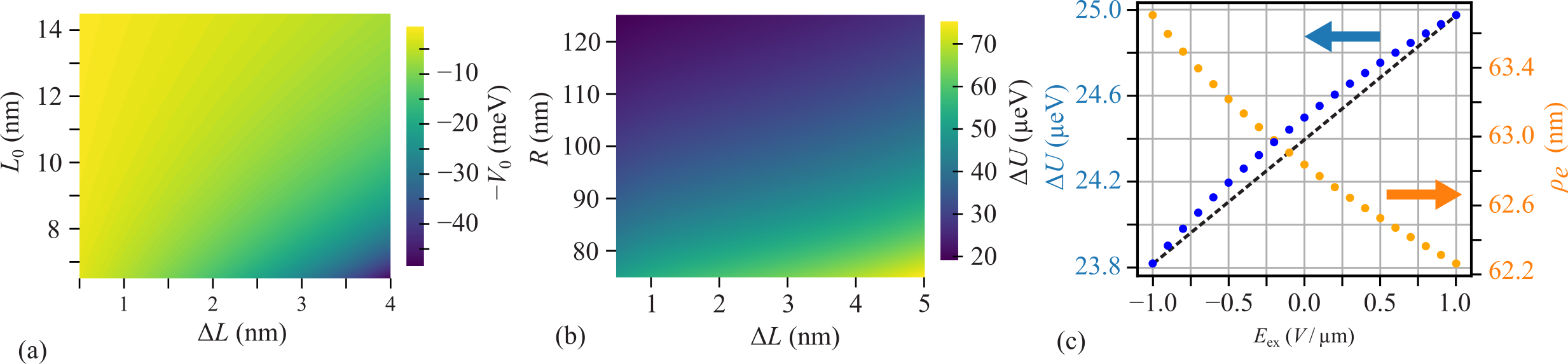}
\caption{ (a) Potential depth $-V_0$ for different $L$ and $\Delta L$. (b)
Lateral excitation energy $\Delta U$ versus $R$ and $\Delta L$ for
$b = 2.0~\mathrm{nm}$ and $E_{\mathrm{ex}} = 0$. (c) $\Delta U$ and mean radius $\protect\rho_e$
as functions of $E_{\mathrm{ex}}$ with $b= 2.0~\mathrm{nm}$
, $R = 110~\mathrm{nm}$, and $\Delta L = 0.5~\mathrm{nm}$.
Black dashed lines indicate linear trends, highlighting the nonlinear
field dependence of $\Delta U$.}
\label{fig:lateral_excite}
\end{figure*}

A uniform external electric field $\vec{E} = E_{\mathrm{ex}}\hat{e}_z$ 
introduces an additional potential:
\begin{eqnarray}
V_{\mathrm{ex}}(z) = 
\begin{cases}
eE_{\mathrm{ex}} \frac{\varepsilon_{0}}{\varepsilon_\mathrm{Ne}}(z+L) & 
\text{for $-L < z < 0$} \\ 
eE_{\mathrm{ex}} \left(\frac{\varepsilon_{0}}{\varepsilon_\mathrm{Ne}} L + z
\right), & \text{for $0 < z$}%
\end{cases}
\label{eq:ExternalElectricField}
\end{eqnarray}
where the zero of potential is defined at the grounded substrate surface ($z = -L$). 
The total potential, $V(z) = V_{\perp}(L, z) + V_{\mathrm{ex}}(L, z)$, enters
the one-dimensional Sch\"{o}dinger equation along the $z$-direction:
\begin{equation}
W(L, E_\mathrm{ex}) \psi = - \frac{\hslash^2}{2m_e} \frac{\partial^2 \psi}{%
\partial z^2}+ \left(V_\perp+ V_\mathrm{ex}\right) \psi,
\end{equation}
where $m_e$ is the electron mass and $\hbar$ is the reduced Planck constant. 
A cutoff distance of $0.23~\mathrm{nm}$ above the Ne surface is introduced 
to account for atomic-scale discreteness~\cite{cole_image-potential-induced_1969,cole_electronic_1971}, i.e., a large repulsive potential barrier of $0.7~\mathrm{eV}$.
Here, the electron's penetration length into the Ne layer is about $0.1~\mathrm{nm}$, and the penetration into the substrate is negligible.
Figure~\ref{fig:perpendicular}(b) shows the calculated ground-state energy $W^{\mathrm{G}}$ 
as a function of layer thickness $L$ for various external electric fields $E_{\mathrm{ex}}$, 
revealing that $W^{\mathrm{G}}$ decreases markedly for thinner layers. For $L=10$ nm, 
$W^{\mathrm{G}}=-44.6~\mathrm{meV}$, which is nearly three times lower than
 the bulk solid Ne value of $-$15.7 meV, indicating a pronounced effect arising from finite layer thickness.
In the perpendicular direction, the excitation energy from the ground to the 
first excited state is approximately 21.1 meV, substantially higher than 
both typical experimental temperatures $\sim 10~\mathrm{mK}$ and the characteristic lateral excitation 
energies (see Fig.~\ref{fig:lateral_excite}). Consequently, excitations along the $z$-direction 
are effectively frozen out under experimental conditions, and only the ground state 
needs to be considered in subsequent analysis.
Figure~\ref{fig:perpendicular}(c) shows that the mean electron height,
$h_e = \int z |\psi(z)|^2 dz$, exhibits minimal variation across a wide range of external electric fields, 
confirming the robustness of perpendicular confinement under typical experimental conditions.

\begin{figure}[t]
\centering
\includegraphics[width=0.75\linewidth]{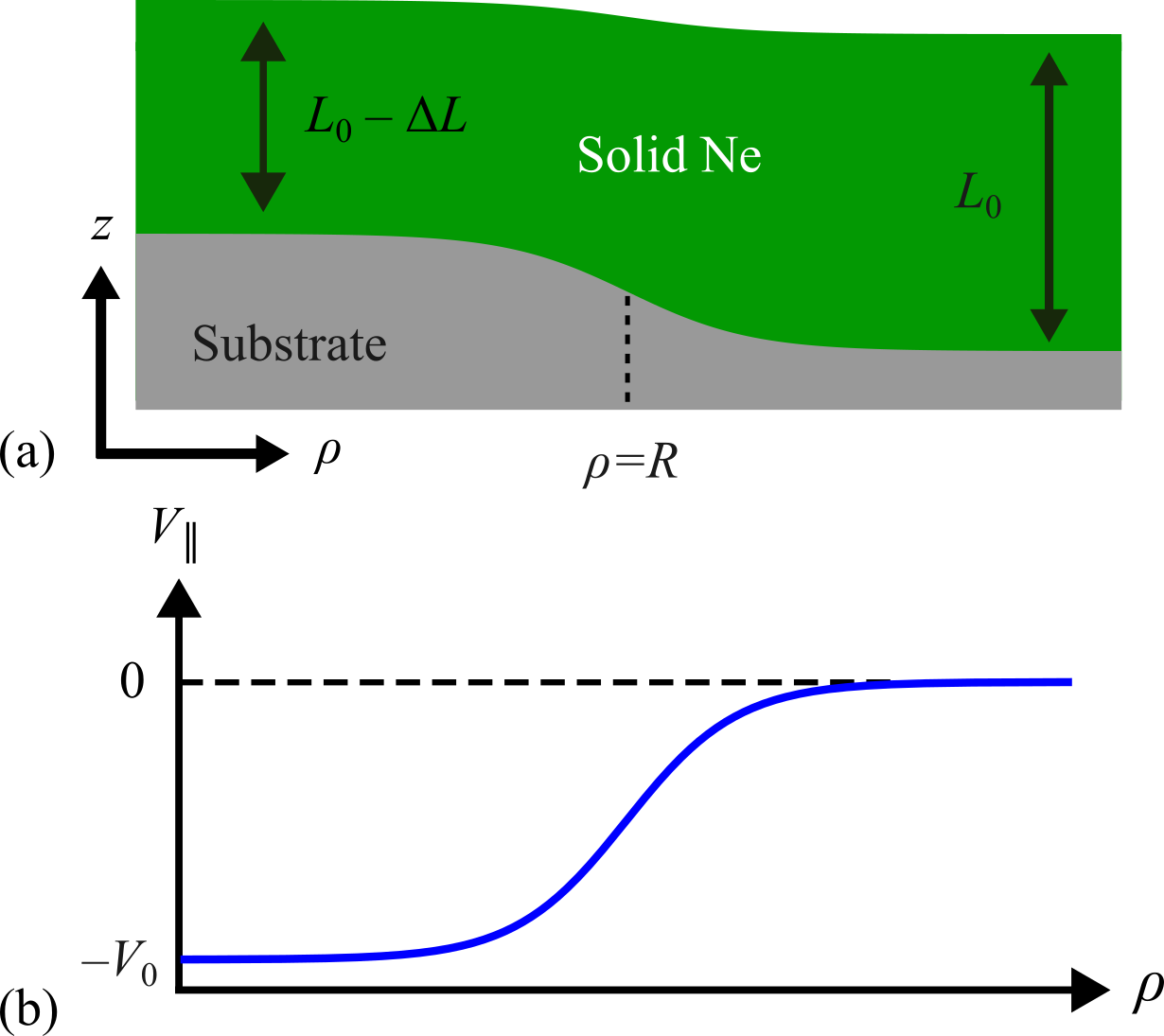}
\caption{(a) Schematic illustration of the model showing the variation in solid Ne layer thickness at $\protect\rho = R$, corresponding to the edge of the substrate nanopillar. (b) Profile of the resulting effective
lateral potential, which varies from $-V_0$ at the center to zero at large
distances.}
\label{fig:2D_schematic_lateral}
\end{figure}

\paragraph*{Lateral trapping from substrate nanopillar geometry.}

As the thickness of the solid Ne layer decreases, the ground-state energy of an 
excess electron correspondingly lowers, indicating that electrons are preferentially
 trapped above thinner regions of the layer surrounded by thicker areas. 
 This observation motivates a lateral trapping mechanism induced by spatial 
 variations in the Ne layer thickness on a substrate nanopillar. During solidification,
 Ne tends to nucleate from the base of a surface step on the nanopillar, 
 leading to local thickness modulation between the pillar top and the surrounding region (Fig.~\ref{fig:2D_schematic}(b)). 
 This spatial variation in the Ne layer thickness can be modeled as
\begin{equation}
L(\rho) = L_0 - \frac{\Delta L}{2} \left(1 - \frac{\rho - R}{\sqrt{(\rho -
R)^2 + b^2}} \right),
\end{equation}
as illustrated in Fig.~\ref{fig:2D_schematic_lateral}(a). Here, the parameters
$\Delta L$, $R$ and $b$ are determined by the substrate nanopillar geometry, while
$L_0 $ is set by the growth and solidification conditions of the Ne layer.

To describe the lateral trapping potential $V_\parallel(\rho, E_{\mathrm{ex}})$, 
we employ a local-thickness approximation (LTA):
\begin{equation}
V_\parallel(\rho, E_{\mathrm{ex}}) = W^G(L(\rho), E_{\mathrm{ex}}) -
W^G(L_0, E_{\mathrm{ex}}),
\end{equation}
which smoothly interpolates between a potential depth of
$-V_0(E_{\mathrm{ex}}) = W^{\mathrm{G}}(L_0 - \Delta L, E_{\mathrm{ex}}) - W^{\mathrm{G}}(L_0, E_{\mathrm{ex}})$ 
near the center ($\rho \ll R$) and zero at large distances ($\rho \gg R$), 
as illustrated in Fig.~\ref{fig:2D_schematic_lateral}(b).
The electric field experienced by the hovering electron originates from the induced surface charge and decays over a characteristic length scale comparable to the electron’s height above the surface. Because the electric field of a point charge falls off as the inverse square of distance, contributions from remote surface charges are negligible. Consequently, the LTA remains valid when the spatial scale of the thickness modulation (i.e., $b$)
is comparable to or larger than the electron height, $h_e \approx 1.7~\mathrm{nm}$ 
for $L_0 = 10~\mathrm{nm}$ (see Fig.~\ref{fig:perpendicular}(c)). 
Moreover, since the binding energy far from the modulation region must match that of a uniform layer and is primarily governed by the local solid-layer thickness, the potential depth is only weakly affected by the LTA assumption. The lateral trapping potential deepens with increasing $\Delta L$ or decreasing $L_0$, as shown in Fig.~\ref{fig:lateral_excite}(a). 

Assuming the cylindrical symmetry of the substrate
nanopillar and the resulting thickness variation, the wave function in the lateral plane
can be expressed as $R_\alpha(\rho)e^{i\alpha\theta}$, and the corresponding 
radial Schr\"{o}dinger equation is given by
\begin{equation}
U_\alpha R_\alpha = -\frac{\hbar^2}{2m_e} \frac{1}{\rho} \frac{d}{d\rho}
\left( \rho \frac{dR_\alpha}{d\rho} \right) + \left( V_\parallel(\rho) + 
\frac{\hbar^2 \alpha^2}{2m_e \rho^2} \right) R_\alpha,
\end{equation}
where $R_\alpha$ is the radial wavefunction, and $\alpha$ is the angular-momentum
quantum number. Figure~\ref{fig:lateral_excite}(b) shows the lateral excitation energy \cite{sup},
$\Delta U = U_1 - U_0$, for different values of $R$ 
and $\Delta L$ under the zero external electric field ($E_{\mathrm{ex}} = 0$), 
with a smoothness parameter $b = 2.0~\mathrm{nm}$. 
The energy levels $U_0$ and $U_1$ define the two states of a single-electron charge qubit.
The excitation energy $\Delta U$ increases as $R$ decreases since it corresponds to the rotational motion of the electron and is primarily determined by the mean radius of the first excited state, 
 $\rho_e = \int \rho |R_1|^2 d\rho$. For deeper potentials, $\rho_e$ remains 
 nearly constant, whereas in shallower potentials the wavefunction extends outward, leading to a larger $\rho_e$ and consequently a smaller excitation energy. Increasing $R$ further promotes wavefunction delocalization, thereby reducing $\Delta U$.

\paragraph{Tuning lateral trapping via external electric field.}

Non-uniformity of the solid Ne layer also introduces spatial variations 
in the electric potential induced by an external field, as described by 
Eq.~\eqref{eq:ExternalElectricField}. In our model, the potential depth 
$V_0$ is approximately shifted by $eE_{\mathrm{ex}}\Delta L\,\varepsilon_0 / \varepsilon_{\mathrm{Ne}}$. 
A positive field ($E_{\mathrm{ex}} > 0$) increases the potential depth, 
leading to modest changes in $\Delta U$, whereas a negative field 
($E_{\mathrm{ex}} < 0$) reduces the depth and leads to a more pronounced shift.
Figure~\ref{fig:lateral_excite}(c) shows $\Delta U$ and $\rho_e$ as functions 
of $E_{\mathrm{ex}}$ for fixed parameters $R = 110~\mathrm{nm}$ and 
$\Delta L = 0.5~\mathrm{nm}$. The results exhibit a clear asymmetric response: $\Delta U$ increases sharply under negative fields and 
more gradually under positive fields. This asymmetry becomes more prominent for sharper thickness transitions (smaller $b$) or shallower potentials, consistent with the observed behavior of $\rho_e$.

To examine whether this nonlinear response persists for smoother substrate 
variations, we consider a model in which the solid Ne layer thickness is described 
by $L(\rho) = L_0(1 + \beta_0 \rho^2)$ with $\beta_0 > 0$, while the upper surface 
of the solid Ne layer remains flat. Assuming that the thickness varies only slightly 
around $L = 10~\mathrm{nm}$ and that the potential depends linearly
on $L$ (see Fig.~\ref{fig:perpendicular}(b)), the lateral trapping potential can be 
approximated as harmonic, $V_\parallel(\rho) \approx \frac{1}{2} m_e \omega_0^2 \rho^2$, 
with a characteristic frequency $\omega_0$. When an external field is applied, the potential becomes 
$V_\parallel(\rho) = \frac{1}{2}
\left( m_e \omega_0^2 + \beta_1 E_{\mathrm{ex}} \right) \rho^2 = \frac{1}{2}
m_e \omega^2 \rho^2$,
where $\beta_1$ quantifies the field-dependent modification of the confinement. The corresponding excitation energy 
$\Delta U = \hbar \omega = \hbar 
\sqrt{\omega_0^2 + \beta_1 E_{\mathrm{ex}}/m_e}$,
naturally reproduces the asymmetric behavior: a sharp increase in $\Delta U$ under negative fields ($-m_e \omega_0^2 / \beta_1 < E_{\mathrm{ex}} < 0$) and a more gradual variation for positive fields.

\paragraph*{Discussion and Conclusion.}

We have shown that the finite thickness of a solid Ne layer has a profound effect on the quantum state and binding energy of an electron floating above its surface. In addition, we have proposed a lateral trapping mechanism arising from local thickness non-uniformities on substrate nanopillars, which may naturally form during the dynamic cooling process in experiments. The corresponding excitation energy is continuously tunable by an external electric field applied along the $z$-direction and exhibits a nonlinear dependence on the field polarity. In the opposite limit, increasing the Ne layer thickness suppresses the lateral trapping effect, thereby enabling the realization of gate-defined qubits, where the electron’s motional states are confined by electrode-induced electric fields, as originally proposed in Ref.~\cite{zhou_single_2022}. 

We further discuss the feasibility of constructing multi-qubit architectures based on this mechanism. In semiconductor systems, nanopillar arrays have been routinely fabricated using lithography and etching techniques, such as in GaAs~\cite{tarucha_shell_1996,steffen_single_1996}. By forming a similar bump and surrounding it with superconducting electrodes, one can create a locally thinner Ne region above the bump to trap an electron, while the electrodes simultaneously serve as a microwave resonator. The electrode height should be slightly lower than that of the central bump to maintain proper field confinement.
For a bump radius of approximately 
 $110~\mathrm{nm}$, comparable to the surface valleys observed 
 experimentally~\cite{zheng_surface-morphology-assisted_2025}, the resulting lateral excitation energy 
is comparable to the typical microwave photon energy ($\sim 26~\mathrm{\mu eV}$)
 used in qubit operations~\cite{zhou_single_2022}.  This correspondence suggests a scalable platform for quantum information processing, where individual qubit sites are periodically patterned on the substrate. Moreover, shaping the bump elliptically can confine the electron’s lateral motion along a preferred axis, reducing the spatial footprint of each qubit and enabling higher qubit density.
 
 Future work may explore coupling between adjacent traps, the role of substrate-induced disorder and decoherence, and the feasibility of entangling operations mediated by shared photonic modes, paving the way toward a solid-Ne-based quantum computing platform.

\begin{acknowledgments}
\emph{Acknowledgment:}
The authors thank Kaiwen Zhang, Kater Murch, and Wei Guo for fruitful discussions and comments.
This work is supported by the Air Force Office of Scientific Research under Grant No. FA9550-20-1-0220 and the National Science Foundation under Grant No. PHY-2409943, OSI-2228725.
\end{acknowledgments}

\widetext
\begin{center}
\textbf{\large Derivation of The Perpendicular Trapping Potential}\label{Appendix}
\end{center}
\setcounter{section}{0}
\setcounter{equation}{0}
\setcounter{figure}{0}
\setcounter{table}{0}
\makeatletter
\renewcommand{\theequation}{A\arabic{equation}}
\renewcommand{\thefigure}{A\arabic{figure}}

Placing an electron at $\vec{r}_0 = (x=0, y=0, z = z_0)$, Poisson's equation is given as 
\begin{equation}
    \nabla \cdot \left\{\varepsilon(r) \nabla\phi(\vec{r})\right\} = -\rho_e(\vec{r}) = e \delta(\vec{r} - \vec{r}_0).
\end{equation}
Introducing the Green function $G(\vec{r}, \vec{r}_0)$ through
\begin{equation}
    \phi(\vec{r}) = \int G(\vec{r}, \vec{r}_d) \rho_e(\vec{r}_d) d\vec{r}_d = -e G(\vec{r}, \vec{r}_0),
\end{equation}
Poisson's equation can be written as
\begin{equation}
    \nabla \cdot \left\{\varepsilon(z) \nabla G(\vec{r}, \vec{r}_0) \right\} = -\delta(\vec{r} - \vec{r}_0),
\end{equation}
and the solution in the cylindrical coordinates $\left(\rho, \theta, z \right)$ is given by
\begin{equation}
G(\vec{r}, \vec{r}_0) = \frac{1}{4 \pi \varepsilon_0} \int_0^\infty J_0(k (\rho - \rho_0)) \left\{ e^{-k|z - z_0|} + \Lambda_L(k) e^{-k(z + z_0)} \right\} dk
\end{equation}
where $J_0$ is the zeroth-order Bessel function and $\Lambda_L$ is the reflection coefficient:
\begin{eqnarray}
\Lambda_L(k) &=& \frac{(\varepsilon_{0} - \varepsilon_{B}) \varepsilon_{Ne} + (\varepsilon_{0} \varepsilon_{B} - \varepsilon_{Ne}^2) \tanh(kL)}{(\varepsilon_{0} + \varepsilon_{B}) \varepsilon_{Ne} + (\varepsilon_{0} \varepsilon_{B} + \varepsilon_{Ne}^2) \tanh(kL)}.
\end{eqnarray}

It is known that the perpendicular trapping potential in the Schr\"{o}dinger equation for the electron floating above the solid neon is well approximated by
\begin{eqnarray}
    V_\perp(\vec{r}) = \frac{-e}{2} \phi(\vec{r}) \nonumber &\approx& \frac{e^2}{2} G(\vec{r}, \vec{r}_0 = \vec{r}) \nonumber \\
    &=& \frac{e^2}{8 \pi \varepsilon_0} \int_0^\infty \left\{ 1 + \Lambda_L(k) e^{-2kz} \right\} dk.
\end{eqnarray}
By subtracting the self-interaction term:
\begin{equation}
    \frac{e^2}{8 \pi \varepsilon_0} \int_0^\infty dk,
\end{equation}
we obtain 
\begin{eqnarray}
    V_\perp(\vec{r}) &=& \frac{e^2}{8 \pi \varepsilon_0} \int_0^\infty \Lambda_L(k) e^{-2kz} dk.
\end{eqnarray}
\end{document}